 \documentstyle[11pt,aaspp]{article}
\newcommand{\BE}{\begin{equation}}
\newcommand{\BEAL}{\begin{eqnarray}}
\newcommand{\EE}{\end{equation}}
\newcommand{\EEAL}{\end{eqnarray}}
\def\_#1{_{\scriptscriptstyle #1}}
\def\&#1{^{\scriptscriptstyle #1}}
\def\Gam#1#2{\Gamma\left({#1\over #2}\right)}
\def\hs{\hat S}

\def\rar{\rightarrow}

\def\deriv#1#2{{d#1\over d#2}}
\def\secder#1#2{{d^2#1\over d#2^2}}
\def\Mg{M_g}

\def\l{\lambda}

\def\n{\nu}
\def\d{\delta}
\def\a{\alpha}
\def\b{\beta}
\def\c{\gamma}
\def\m{\mu}
\def\e{\epsilon}

\def\r{\rho}

\def\ff{\varphi}

\def\t{\tau}
\def\mo{m_0}
\def\ro{r_0}

\def\s{\sigma}
\def\sr{\sigma_r}
\def\st{\sigma_t}
\def\sp{\sigma_\parallel}

\def\hm{\hat m}
\def\hmo{\hat\mo}

\begin{document}
\title{Models for anisotropic spherical stellar systems
with a central point mass and Keplerian-fall-off velocity dispersions}
\author{ Mordehai Milgrom}
\affil{Department of Condensed-Matter
 Physics, Weizmann Institute of Science  76100 Rehovot, Israel}
\begin{abstract}
We add to the lore of spherical, stellar-system models a two-parameter
family with an anisotropic velocity dispersion, and a central point mass
(``black hole'').
The ratio of the tangential to radial dispersions is
constant--and constitutes the first parameter--while each decreases with
radius as $r^{-1/2}$. The second parameter is the ratio of the central
point mass to the total mass. The Jeans equation is solved to give the
density law in closed form: $\r\propto (r/\ro)^{-\c}/
[1+(r/\ro)^{3-\c}]^2$, where $\ro$ is an arbitrary scale factor. The two
parameters enter the density law only through their combination $\c$.
At the suggestion of Tremaine, we also explore models with only the
root-sum-square of the velocities
 having a Keplerian run, but with a variable anisotropy
ratio. This gives rise to a more versatile class of models, with analytic
expressions for the density law and the dispersion runs, which contain
more than one radius-scale parameter.
\end{abstract}
\section{INTRODUCTION}
\par
Model spherical stellar systems are useful in various contexts, as
 models of elliptical galaxies, star clusters, cores of galaxies, etc.,
 and also as test beds for various theoretical ideas.
To the library of classic models such as the Emden polytropes,
 isothermal spheres, King, and Michie models [see e.g. \cite{bt87}
and references therein],
 were added, in recent years, the models of \cite{jaf83}, of \cite{bs89},
and of \cite{her90}
 and their generalizations to ``eta'' models by \cite{deh93}, and by
\cite{tre94}.
\par
There are various routes to constructing such spherical models:
One may start from a distribution function (hereafter, DF) satisfying the
 Jeans theorem, and see if one gets a useful density and dispersion
 profiles. One is then ensured from the start that the model is an
 equilibrium one (but stability may still be a question).
Trial and error is then required to obtain useful models, with
relevant density and velocity-dispersion profiles.
Alternatively, we may start by dictating constraints on the more directly
observable density, $\r(r)$, and the runs of tangential and radial
velocity dispersions,  $\st(r)$ and $\sr(r)$, respectively.
 A necessary (but not sufficient)
condition that there be an underlying stationary DF
is that these three functions satisfy the Jeans equation
\BE {1\over\r}\deriv{(\sr^2\r)}{r}+{2\b\sr^2\over r}=-{M(r)G\over r^2},
 \label{jeansi} \EE
with
\BE M(r)=4\pi\int_0^r r'^2~\r(r')~dr',   \label{masar} \EE
and
\BE \b\equiv 1-{\st^2\over \sr^2}. \label{betat} \EE
This is still only one functional constraint on the three functions,
so to pinpoint a model we need two further functional
 constraint. For example,
we can restrict ourselves to isotropic models and put
$\sr(r)=\st(r)=\s(r)$.
Then we can start with a density profile and calculate $\s(r)$ from
the Jeans equation, as in \cite{tre94}.
We can, alternatively,
assume an equation of state--a relation between $\r$ and $\s$--and solve
 the Jeans equation for both $\r(r)$ and $\s(r)$ (as for gas
 polytropes).
Or, we may start with an assumed $\s(r)$ and solve the Jeans equation
 for $\r(r)$ (as in the case of isothermal spheres).
\par
If we want anisotropic models we may start with a different constraint
on $\sr(r),~\st(r)$. For example, we may assume that their ratio
is position independent (as we do in much of this paper),
 and then dictate
either a density or a dispersion profile.
\par
Such models are useful if at least
some aspects of them can be given in simple, closed forms.
In the end one has to ascertain that the model is underpinned by
a positive DF satisfying the Jeans theorem, if one seeks to insure
that the model is a realization of at least one equilibrium system.
\par
We offer here what we believe is a novel family of such models.
We start by dictating the run of velocity dispersions: a constant
 ratio between the
tangential and radial dispersions, while each of them decreases,
in a Keplerian manner, as $r^{-1/2}$.
It is clear that in order to obtain a bound system with a finite total
mass the dispersions must decline at least that fast asymptotically
(lest they outstrip the escape velocity, which decreases in this manner).
In fact, for a fixed-$\b$ the dispersions must behave so asymptotically.
Such a behavior is also ``natural'' for a stellar system near a central,
dominant point mass.
We then extend this family of models by assuming that only
$(\sr^2+\st^2)^{1/2}$ is exactly Keplerian, but the anisotropy ratio
is not constant.
\par
 The useful features of the models are: simple, closed-form
 expressions for the density, accumulated mass, velocity dispersions,
and potential, and the effortless inclusion of a central point mass.
\par
We describe the constant-$\b$
 models in section 2, derive some additional properties
 in section 3 (including a partial discussion of underlying
DFs), and explore a generalization in section 4.
\section{THE MODELS}
\par
We substitute our assumed dispersion profile,
\BE  (1-\b)^{-1}\st^2=\sr^2=S/r,  \label{lawa} \EE
in the Jeans equation(\ref{jeansi}) to obtain
\BE \deriv{ln~\r}{ln~r}+2\b-1 =-m(r), \label{hata} \EE
where, $m(r)\equiv M(r)G/S$ is the dimensionless accumulated mass
($\hat M\equiv SG^{-1}$ has the dimensions of mass and can be used to
normalize all the masses in the problem), and $\b$ is constant.
\par
The addition of a point mass $M_0$ at the center is effected by
 subtracting from the right-hand side of eq.(\ref{hata}) the constant
\BE  \mo\equiv M_0/\hat M= M_0G/S.  \label{mosa} \EE
We then obtain for the Jeans equation
\BE \deriv{ln~\r}{ln~r}+\c =-m(r), \label{hapsss} \EE
where
\BE \c\equiv 2\b+\mo-1, \label{gagat} \EE
and $m(r)$ includes only the accumulated mass of the stellar ``gas'',
excluding the point mass; we thus have $m(0)=0$.
Equation(\ref{hapsss}) can be solved analytically:
We eliminate $\r$ using
\BE \r={\hat M \over 4\pi r^2}m',  \label{lauar} \EE
to get an equation for $m(r)$ ($m'$ is the $r$-derivative of $m$)
\BE rm''+(\c-2)m'+mm'=0. \label{yiyiu} \EE
The left-hand side is the derivative of $rm'+(\c-3)m+m^2/2$, which must
then be a constant. Since $m(0)=0$, this constant must be 0
so we get
\BE m^{-1}(3-\c-{1\over 2} m)^{-1}m'={1 \over r}.
 \label{xcexx} \EE
 For $\c\ge 3$, $m'$ is negative, which is unphysical.
 For $\c< 3$, the equation is
straightforwardly integrated to yield the desired mass profile
\BE m(r)=2(3-\c){x^{3-\c}\over 1+ x^{3-\c}},  \label{olkse}\EE
where $x=r/\ro$, and the scale radius, $\ro$,
 is an arbitrary integration constant; it equals the half-mass radius.
 \par
This mass profile corresponds to the density profile
\BE  \r(r)={(3-\c)^2\hat M\over 2\pi \ro^3}{x^{-\c}\over
(1+x^{3-\c})^2}.  \label{tic} \EE
Near the origin, the density behaves as $r^{-\c}$, while at infinity it
 behaves as $r^{\c-6}$.
\par
We see from eq.(\ref{olkse}) that the total dimensionless mass is
 $m(\infty)= 2(3-\c)$, or
\BE G\Mg=2(3-\c)S=4(2-\b)S-2M_0G.  \label{lmaoe} \EE
$S$ is thus determined by the two masses: the total ``gas'' mass, $\Mg$,
 and the central point mass, $M_0$, through
\BE  S=[4(2-\b)]^{-1}G(\Mg+2M_0).   \label{pyooot} \EE
This is the virial relation for the models.
We shall see in section 3
 that for $\c\ge 2$ the kinetic and potential energies
of the model are infinite, but relation(\ref{pyooot}) still holds for
all legitimate values of $\c< 3$, and follows directly from the
structure equation.
\par
The two masses determine the parameter $\mo$ through
$\mo=4(2-\b)M_0/(\Mg+2M_0)$, and  $\c$ through
\BE \c=2\b-1+4(2-\b){M_0/\Mg\over 1+2M_0/\Mg}. \label{yyyta} \EE
\par
In the limit of test-particle ``gas'', $\Mg\rar 0$, with $\b$ fixed,
 we have  $\c\rar 3$,
 so the model becomes meaningless. A meaningful test-particle model is
 obtained if we let $\b\rar -\infty$ (circular orbits)
such that $\c<3$; $\b\Mg$ is then fixed at $(\c-3)M_0/2$.
Then $S\rar 0$, but $2\st^2r\rar -2\b S\rar GM_0$, and we get a model
with test particles on Keplerian
 circular orbits around a point mass, the likes
of which can be built with any density distribution. If we admix
test particles on non-circular orbits the Keplerian fall-off which
we require cannot be maintained.
\par
In the limit $\b\rar -\infty$, but $\Mg$ and $M_0$ kept constant, we have
$\c\rar -\infty$, and the sphere becomes an infinitely thin shell at
an arbitrary radius $\ro$, with the particles moving on circular
orbits: A self gravitating sphere cannot consist of particles on
circular orbits with Keplerian velocities unless they are all at the same
radius.
\par
 For $\c=2$, one obtains the density distribution of the Jaffe
model [\cite{jaf83}] [which is the same as the ``eta'' model of
\cite{tre94}, and \cite{deh93} with $\eta=1$].
 This value of $\c$ can be obtained only with a
 central mass as the maximum value of $\c$ for $\mo=0$ is 1. In the
isotropic case it corresponds to $M_0/\Mg=3/2$.

\section{GENERAL PROPERTIES}
{\bf Energies}
\par
We now consider the potential run of the model.
The point mass makes the usual contribution to the potential;
that of the stellar ``gas'', $\ff_g$,
 can be straightforwardly integrated from the
expression
\BE \deriv{\ff_g}{r}=-{GM(r)\over r^2}=-{G\Mg\over\ro^2}{x^{1-\c}\over
1+x^{3-\c}}  \label{lulio} \EE
to give
\BE \ff_g(r)=-{G\Mg\over r}~{}_2F_1[1,(3-\c)^{-1};
(4-\c)(3-\c)^{-1};-(\ro/ r)^{(3-\c)}], \label{hyper} \EE
where ${}_2F_1$ is the hypergeometric function [see e.g. \cite{gr80}].
As  ${}_2F_1(a,b;c;0)=1$  we have asymptotically
$\ff_g(r)\rar -{G\Mg\over r}$, as expected.
\par
As to the behavior of $\ff_g$ at the origin, we use the behavior of
${}_2F_1$ at large values of its argument to deduce that for $\c<2$
\BE \ff_g(0)=-{G\Mg\over \ro}\Gam{2-\c}{3-\c}
\Gam{4-\c}{3-\c}.  \label{kasasa} \EE
 For $\c=2$, $\ff_g$ diverges logarithmically at the origin, and for
$2<\c<3$, $\ff_g$ diverges as $r^{2-\c}$.

Some special cases: For $\c=1$, one needs
 ${}_2F_1(1,1/2;3/2;-x^{-2})=x~tg^{-1}(1/x)$ to get
\BE\ff_g(r)=
-{G\Mg\over \ro}tg^{-1}\left({\ro \over r}\right). \label{fff} \EE
 For $\c=2$,
 ${}_2F_1(1,1;2;-x^{-1})=x~ln(1+1/x)$, yielding the known expression
for the Jaffe model $\ff_g(r)=(G\Mg/\ro)ln[r/(r+\ro)]$.
\par
The total kinetic energy of a model sphere is infinite for $\c\ge 2$, and
for $\c<2$ is
\BE E_k={S\Mg\over\ro}{3-2\b\over 2}\Gam{2-\c}{3-\c}\Gam{4-\c}{3-\c}.
\label{paaaui} \EE
The potential energy of the stellar ``gas'' is, for $\c<2$,
\BE E_p=-{G\Mg^2\over 2\ro}\Gam{2-\c}{3-\c}\Gam{4-\c}{3-\c}
\left({2-\c\over3-\c}+{2M_0\over \Mg}\right). \label{plalai} \EE
By eq.(\ref{pyooot}) they satisfy $E_k=-E_p/2$.

{\bf Projected properties}
\par
If we define the integrals
\BE I(\m,\n;\l)\equiv \int_{\l}^{\infty}{x^{-\m}(x^2-\l^2)^{-1/2}
\over (1+x^{\n})^2}dx,  \label{intata} \EE
then the projected surface density $\Sigma(a)$, at a projected radius
 $a=\l\ro$, is given by
\BE \Sigma(a)={3-\c\over 2\pi}{\Mg\over\ro^2}I(\c-1,3-\c;\l).
\label{bububu} \EE
The line-of-sight (projected) velocity dispersion, $\sp(a)$, is
\BE \sp^2(a)={S\over \ro}{I(\c,3-\c;\l)-\b\l^2I(\c+2,3-\c;\l)\over
I(\c-1,3-\c;\l)}.  \label{sisisi} \EE
At large projected radii ($a\gg\ro$) we can write
\BE \Sigma(a)\approx
{3-\c\over 4\pi^{1/2}}{\Mg\over\ro^2}{\Gam{5-\c}{2}\over
\Gam{6-\c}{2}}\l^{\c-5},  \label{cucucu} \EE
and the line-of-sight dispersion
\BE \sp^2(a)\approx
{\Gamma^2\left({6-\c\over 2}\right)
\over \Gam{7-\c}{2}\Gam{5-\c}{2}}\left(1-\b{6-\c\over 7-\c}\right)
{S\over a}.
\label{laflaf} \EE
As examples, for $\c=1$ we get asymptotically
\BE \sp^2(a)\approx {9\pi\over 32}(1-{5\over 6}\b){S\over a}, \label{pas}
\EE
 for $\c=2$
\BE \sp^2(a)\approx {8\over 3\pi}(1-{4\over 5}\b){S\over a}, \label{pip}
\EE
 and for $\c\approx 3$
\BE \sp^2(a)\approx {\pi\over 4}(1-{3\over 4}\b){S\over a},\label{baypip}
\EE
Near the origin ($a\ll\ro$) we can write, when $\c>1$,
\BE \Sigma(a)\approx
{3-\c\over 4\pi^{1/2}}{\Mg\over\ro^2}{\Gam{\c-1}{2}\over
\Gam{\c}{2}}\l^{1-\c},  \label{dududu} \EE
and
\BE \sp^2(a)\approx
{\Gamma^2\left({\c\over 2}\right)
\over \Gam{\c+1}{2}\Gam{\c-1}{2}}\left(1-\b{\c\over \c+1}\right)
{S\over a}. \label{zamzam} \EE
 For example, for $\c=2$
\BE \sp^2(a)\approx
{2\over \pi}(1-{2\over 3}\b){S\over a}. \label{mam}
\EE
 For $\c\approx 3$, the behaviors near the origin and at large radii are
the same.
We see that unlike the density, and surface-density, run,
the projected-dispersion run depends on both $\b$ and $\c$.
 For $\c>1$, $\sp$ declines as $r^{-1/2}$ at both ends; the
 proportionality constant is, in general, different, but may be the same
for special choices of $\b$ and $\c$ (for example, for $\c=2$, $\b=5/6$).

{\bf Underlying distribution functions}
\par
We have not been able to ascertain that all the models discussed here
have underlying non-negative DFs that satisfy the Jeans theorem.
 For the isotropic case, $\b=0$, one can try to search for a DF that
 depends on the particle energy alone. The procedure is then
 straightforward.
 First, we note from eq.(4-139) in \cite{bt87} that for this limited
class of DFs the density must not increase with radius anywhere.
This necessary condition excludes the (isotropic)
 models with $\c<0$, for
 which $\r$ increases near the origin.
The DF is uniquely determined by the density distribution
through the Eddington relation [eq.(4-140b) of \cite{bt87}]
\BE f(\e)={1\over 8^{1/2}\pi^2}\left[\int_0^{\e}\secder{\r}{\Psi}
{d\Psi\over\sqrt{\e-\Psi}}+{1\over\sqrt{\e}}\left(\deriv{\r}{\Psi}
\right)\_{\Psi=0}\right]. \label{lmalma} \EE
Here, $\e=-E$, where $E$ is the energy, $\Psi=-\ff$, and $\r$ is
viewed as a function of $\Psi$ as both are functions of $r$.
The value $\Psi=0$ occurs at radial infinity, where
$\r\propto r^{\c-6}$ decreases faster than $\Psi\propto r^{-1}$;
thus, the second term in eq.(\ref{lmalma}) vanishes.
If we change the integration variable to $r$,
eq.(\ref{lmalma}) becomes
\BE
 f(\e)={1\over 8^{1/2}\pi}\int_{\Psi^{-1}(\e)}^{\infty}
-\secder{\r}{\Psi}\deriv{\Psi}{r}{dr \over\sqrt{\e-\Psi}}.
\label{tutoa} \EE
One can show that
\BE -\secder{\r}{\Psi}\deriv{\Psi}{r}=\r{m+\c\over m+\c+1}\left[\c-1+
m{m^2+m(2\c+3/2)+(\c-1)(\c+3)\over (m+\c)(m+\c+1)}\right]. \label{ppa}\EE
Thus, for $\c\ge 1$ the integrand is non-negative, and the DF is always
positive. For $\c<1$, the integrand becomes negative for small enough
radii, and the DF becomes negative for high enough energies.
Such models do not have legitimate DFs that depend only on $E$.
\par
The two limiting models discussed at the end of section 2 clearly also
have legitimate underlying DFs.

\section{A GENERALIZATION}
\par
Scott Tremaine (private communication) has pointed out to us that the
crucial step in our derivation, of once integrating the Jeans equation
[going from eq.(\ref{yiyiu}) to eq.(\ref{xcexx})], does not require that
$\sr$ and $\st$ are separately
 Keplerian, it is enough that
$(\st^2+\sr^2)^{1/2}$ is. Making only this relaxed assumption permits
us to explore a larger family of models, which we now proceed to do.
\par
The general Jeans equation(\ref{jeansi}), in the presence of a central
 point mass $M_0$, can be brought to the form
\BE (r^2M'\sr^2)'-[2r(\sr^2+\st^2)-GM_0]M'+{G\over 2}(M^2)'=0,
\label{patuna} \EE
where the derivation is with respect to $r$.
Now assume only that
\BE \sr^2+\st^2=\hs r^{-1}, \label{lmlmlm} \EE
and, as before, define the dimensionless quantities
 $\hmo\equiv GM_0/\hs$,
$\hm(r)\equiv GM(r)/\hs$, and also
\BE s(r)\equiv \sr^2(r)r/\hs={1\over 2-\b(r)}. \label{fievsy} \EE
Then, the Jeans equation becomes
\BE (rs\hm')'-\eta \hm'+{1 \over 2}(\hm^2)'=0,  \label{juijui} \EE
where $\eta\equiv 2-\hmo$.
As explained in section 1, having imposed only one functional
condition on the three functions $\r,~\sr$, and $\st$, we have yet to
specify another in order to pinpoint a model. This
 can be a dictation of the run of the anisotropy parameter, as we
shall do below by dictating $s(r)$.
 Note that $0\le s(r)\le 1$, where $s=0,1$ for purely tangential,
and purely radial dispersion, respectively.
\par
As before, the left-hand side of eq.(\ref{juijui}) is a derivative, and
is readily integrated once. As $s$ is bounded, the integration constant
vanishes, and we get
\BE rs\hm'-\eta \hm+{1 \over 2}\hm^2=0.  \label{jahols} \EE
Values of $\eta\le 0$ are non-physical as they give a negative $m'$.
 For $\eta>0$, the integration of eq.(\ref{jahols}) gives the mass
 profile
\BE \hm(r)={2\eta e^{\eta X(r)} \over 1+e^{\eta X(r)}}, \label{sbuu} \EE
where
\BE X(r)=\int_{\ro}^{r} {dr\over rs(r)}, \label{huhui} \EE
and $\ro$ is some arbitrary radius that will appear as a scale in the
density law. The integration constant in eq.(\ref{huhui}) is chosen
so that $\ro$ is the half-mass radius.
Since $s$ cannot exceed unity, $X(r)$ must diverge (at least
 logarithmically) for $r\rar \infty$, and so $\hm(\infty)=2\eta$,
yielding the virial relation
\BE 4\hs=G(\Mg+2M_0).  \label{atuion} \EE
Thus, $\eta$ is determined by
\BE \eta={2\Mg\over \Mg+2M_0}.  \label{lampadae} \EE
The results for the constant-$\b$ case are reproduced when we note that
in this case $S$ and $\hs$ are related by
$\hs=(2-\b)S$, so $\hm=m/(2-\b)$, $\eta=(3-\c)/(2-\b)$, etc..
\par
The resulting density run is
\BE \r(r)={\Mg\over 2\pi r^3 s(r)}{\eta e^{\eta X}\over
(1+e^{\eta X})^2}.   \label{rastae} \EE
At large $r$, where $X(r)$ is positive and diverges, the asymptotic
form of the density is
\BE \r(r\rar\infty)\approx{\eta\Mg\over 2\pi r^3 s(r)} e^{-\eta X};
   \label{palgae} \EE
near the origin, where $X(r)$ also diverges, but is negative,
\BE \r(r\rar 0)\approx{\eta\Mg\over 2\pi r^3 s(r)} e^{\eta X}.
   \label{saplae} \EE
\par
One can try different forms of $s(r)$, subject to $0\le s(r)\le 1$,
and see if any interesting mass distributions result.
Some general points first:
If we depart from the constant-$\b$ models discussed above,
a new radius scale must be introduced by the choice of $s$, beside the
scale, $\ro$, introduced through the integration constant. This is
 because $s$ must be of the form $s(r/a)$.
\par
Near the origin the behavior is like that of the constant-$\b$
models with $\b=\b(0)$. Thus, if $s(0)\not =0$ [$\b(0)\not =-\infty$],
 $X(r)$ diverges logarithmically (and is negative) at
 the origin, and we get from eqs.(\ref{huhui})(\ref{saplae})
a power-law density behavior: $\r\propto r^{-[3-\eta/s(0)]}$.
If $s(0)=0$, $X(r)$ diverges faster than logarithmically, and $\r$
 vanishes non-analytically there [as in the constant $s(r)=0$ case,
where the ``gas'' is concentrated in an infinitely thin shell].
\par
At large radii, if $s(\infty)\not = 0$, $X$ diverges logarithmically,
 and $\r$ has a power-law behavior:
 $\r\propto r^{-[3+\eta/s(\infty)]}$.
 If $s(\infty)=0$, $X$
diverges faster, and $\r$ vanishes faster than a power (e.g. as an
exponential of a power).
\par
Of the many models that can be produced we consider one class
 in more detail:
Take $s(r)$ to vary monotonically between the values $\n$ at $r=0$,
 and $\t$ at infinity as
\BE s(r)={\n+\t (r/a)^d\over 1+(r/a)^d}.  \label{nutat} \EE
Here, $d$ is a power that controls the sharpness of the
 transition of $s$ from $\n$ to $\t$, and, when either
vanishes, $d$ measures the speed with which $s$ does ($d\rar -d$ is
 tantamount to $\n\leftrightarrow\t$).
We take $\n\not =0$ ($\n=0$ can be treated by changing the sign of $d$),
then, for $\t\not =0$, the integral in expression(\ref{huhui}) for $X$
gives
\BE e^{\eta X}={x^{\a}(\n+\t x^d)^{\d/d}\over
x_0^{\a}(\n+\t x_0^d)^{\d/d}},  \label{buerqa} \EE
where $x\equiv r/a$, $x_0\equiv \ro/a$, $\a\equiv \eta/\n$, and
$\d\equiv \eta/\t-\eta/\n$.
The mass profile is thus
\BE \hm(r)={2\eta x^{\a}(\n+\t x^d)^{\d/d}\over
 x_0^{\a}(\n+\t x_0^d)^{\d/d}+
 x^{\a}(\n+\t x^d)^{\d/d}},  \label{mayutra} \EE
and evinces the expected appearance of two distinct scale lengths.
The density law is obtained by substituting expression(\ref{buerqa}) in
eq.(\ref{rastae}):
\BE \r(r)={\eta\Mg\over 2\pi r^3 s(r)}
{ x^{\a}(\n+\t x^d)^{\d/d}  x_0^{\a}(\n+\t x_0^d)^{\d/d}\over
[x_0^{\a}(\n+\t x_0^d)^{\d/d}+
 x^{\a}(\n+\t x^d)^{\d/d}]^2}.    \label{lagrae} \EE
 The density is a power law in radius at both ends
with a power $-(3-\eta/\n)$ at the origin, and $-(3+\eta/\t)$
 at infinity.
\par
The case $\t=0$ requires special treatment: here one finds
\BE e^{\eta X}=\left({x\over x_0}\right)^{\a}
e^{(\a/d)(x^d-x_0^d)}, \label{gaty}
\EE
giving rise to a density run that decreases exponentially at large $r$:
\BE \r(r)\propto r^{-(2+\a)}e^{-(\a/d) (r/a)^d},  \label{jiaerq} \EE
for $r\rar\infty$.
\par
We thus get a class of models that are more flexible as regards the
density run, but they contain two scale lengths instead of one, and
are certainly less wieldy.

\acknowledgements
I thank Scott Tremaine for useful comments  and suggestions.

%\clearpage

%\clearpage***
\end{document}